\documentclass{emulateapj}
\usepackage{times}
\usepackage{graphicx}% Include figure files

\renewcommand{\arraystretch}{0.9}

\slugcomment{To appear in ApJ}

\shorttitle{$\gamma$-ray Pulsar Light Curves}
\shortauthors{}

\begin{document}

\title{Constraining Pulsar Magnetosphere Geometry with ${\gamma}$-Ray Light Curves}

\author{Roger W. Romani\altaffilmark{1} and Kyle P. Watters\altaffilmark{1}}
\altaffiltext{1}{Department of Physics, Stanford University, Stanford, CA 94305}
\email{rwr@astro.stanford.edu, kwatters@stanford.edu }

\begin{abstract}

	We demonstrate a method for quantitatively comparing
$\gamma$-ray pulsar light curves with magnetosphere beaming models.
With the {\it Fermi} LAT providing many pulsar discoveries and
high quality pulsar light curves for the brighter objects,
such comparison allows greatly improved 
constraints on the emission zone geometry and the
magnetospheric physics. Here we apply the method to {\it Fermi}
LAT light curves of a set of bright pulsars known since 
{\it EGRET} or before. We test three approximate models for the
magnetosphere structure and two popular schemes for the location
of the emission zone, the Two Pole Caustic (TPC) model and
the Outer Gap (OG) model. We find that OG models and relatively
physical $B$ fields approximating force-free dipole magnetospheres
are preferred at high statistical significance. An application
to the full LAT pulsar sample will allow us to follow the 
emission zone's evolution with pulsar spindown.
\end{abstract}


\section{Introduction}

	With the successful launch of {\it Fermi}, formerly {\it GLAST},
the number and quality of $\gamma$-ray light curves available for comparison
with magnetospheric models has substantially improved \citep{psrcat}. 
%In particular, the first serious sample of $\gamma$-ray selected pulsars
%have been assembled, which avoids classical radio selection bias.
The quality and uniformity of the data make this a good time to
down-select models for pulsed magnetospheric emission, by comparison with the
observed curves. In the present paper we describe a method of attack for
this process and apply it to the new improved light curves provided
by {\it Fermi} for several well-known $\gamma$-ray pulsars.

	The most popular pulsar models postulate regions in the
magnetosphere where the force-free perfect MHD (${\vec E}\cdot {\vec B}=0$) 
conditions break down, and in response to 
the uncanceled rotation-induced EMF ${\vec E}\approx -{\vec r} \times {\vec \Omega}
\times {\vec B}/c$, the charges re-arrange so that non-negligible fields
penetrate these lower density 'gap' zones, causing particle acceleration and radiation.
Traditionally, one approximates the magnetosphere by a vacuum dipole field.

There are three locations commonly discussed for these magnetospheric gaps.
The earliest pulsar models focused on acceleration at the foot points
of the open field line zone, the so-called ``polar caps''. This emission
should arise from within about a stellar radius \citep{dh96}, and should 
therefore suffer strong
attenuation from one photon ($\gamma$-B) pair creation. The corresponding
hyper-exponential cutoffs are not seen in the LAT pulsar spectra, and so it
has been concluded that the bulk of the pulsar emission must
come from higher altitudes \citep{psrcat}. The first high altitude model posits 
``Holloway'' gaps above the null charge surface,
extending toward the light cylinder at $r_{LC}=c/\Omega$
(the ``outer gap'' model, Cheng, Ho \& Ruderman 1986; Romani 1996).
More recently, it has been argued that acceleration at the rims of 
the polar caps may extend to very high altitude (Muslimov and Harding 2004), 
effectively pushing the polar cap activity outward. In this `slot gap'-type
picture emission can occur over a large fraction of the boundary of the 
open zone.  The geometrical realization of this model makes low altitude 
emission visible from one
hemisphere and higher altitude ($r>r_{NC}$) emission visible from
the other. This is the ``two pole caustic'' (TPC) model (Dyks and Rudak 2003)
which is thus intermediate between the polar cap (PC) and outer gap (OG) pictures.

	In Watters et al. (2009), we created a library of light curves for these
three emission locations in a dipole field and classified the phases of the dominant peaks. The
results were presented as an `atlas' of pulse properties which could be used
to predict the pulse multiplicity and phase separation of a given model for a
particular viewing geometry. While this is useful for an overview of the
possible light curves and provides a convenient reference to read off the
allowed angles for a given model, the complexity of the modeled beams 
meant that an appropriate phase separation could be found in several models
for any one pulsar.

	These computations implicitly assume static co-rotating charges
only.  Of course, the presence of radiating pair plasma in the magnetosphere
requires some particle production, and so the vacuum plus co-rotating charge 
model will be approximate at best. Therefore an alternative attack is
based on numerical solutions for magnetospheric charges, currents and
fields, assuming that pair production is so robust that the force-free
condition holds everywhere \citep{cet99,spit06,bs09b}. Such filled magnetospheres
lack the acceleration fields required to produce powerful $\gamma$-ray beams, 
so the truth must lie in between,
with some regions experiencing charge starvation and departing from force-free.

	Here we wish to show how the actual light curves
from the computations can be compared with the data.  We demonstrate
that the present data quality are sufficient to start comparing predictions
between different assumed magnetic field structures and emission zone
locations. In particular, while in many cases several models can produce
plausible matches to the light curves for {\it some} angle, when we
restrict our attention to the geometries demanded by external constraints,
one can often strongly exclude certain models.

	For our model set, we start by amending certain inadequacies in the vacuum
computations pointed out by \citet{bs09a}. This ends up making modest
differences to the results in \citet{wet09}. 
We also introduce a simple model to illustrate the field structure 
perturbations resulting from open-zone currents. Comparing with the
{\it Fermi} full band light curves, we find that lower altitude TPC-type
models have great difficulty producing acceptable matches for many pulsars
for all of these field geometries (at least for the assumptions made here)
and that OG-type models are statistically strongly preferred. Further,
in {\it Fermi}-quality data we find that the light curve matches can
often be significantly improved by the magnetospheric current perturbations.
In a following communication, we will apply such analysis to the full set
of {\it Fermi} LAT pulsars, marry the light curve fitting code to
a population synthesis model, and extract additional information on the pulsar 
beam structure and its evolution with age as the pulsar spins down.
	
\section{Modeling Method}

	Our basic assumptions are that $i$) the stable, time-averaged
pulse profile is caused by emission zones locked to a set of 'open' 
field lines, $ii$) these field structures are dominated by
the dipole component of the field anchored in the star, since for young 
pulsars this emission occurs at many $R_\ast$, $iii$) that a co-rotation
charge density distributed throughout the bulk of the magnetosphere
causes the plasma to follow the field lines and rotate with the star
and $iv$) in the charge-starved gap zones the uncanceled field rapidly accelerates
$e+/e-$ to highly relativistic energies so that these $\gamma$-ray producing
particles also follow the magnetic field lines.

	As has been well established, starting with Morini (1983) and 
discussed explicitly by Romani \& Yadigaroglu (1995), the combination
of field line sweep back, co-rotation of the radiating particles
and travel time across the light cylinder tends to pile 
up the emission into 'caustics' in pulsar phase. As noted by
\citet{bs09a} this reflects the singularity of the Jacobian
matrix mapping from the emission surface to the antenna pattern on the sky.
The singularity naturally
produces sharp pulses with emission from a range of magnetospheric
altitudes, a property shared by all competitive magnetospheric emission
models. The sharpness of the peaks of the pulsed emission \citep{vela1}
implies that the `width' of the radiating zone is modest, at least in 
a time averaged sense. However, the resolved rise to peaks of the
observed $\gamma$-ray pulses suggest either a wide range of emission height,
a fundamental radiation pattern with wings, or a dispersion in
the direction of the emission zone, again in a time-averaged sense.

	Once one chooses a prescription for the magnetic field structure,
one can follow radiation from the active emission zone to a $(\phi, \zeta)$ 
observer plane, where $\phi$ is the pulsar spin phase, with phase zero 
when the surface magnetic dipole axis passes closest to the observer 
line of sight. The observer line of sight is at viewing angle $\zeta$ 
to the spin axis.  This two dimensional intensity map is the `antenna pattern'
for the pulsar radiation \citep{wet09}. Once one specifies the inclination
$\alpha$ of the magnetic dipole axis to the rotation axis one has
a complete geometrical description. Of course, the choice
of the radiating zone and the (energy-dependent) intensity weight
through that emission zone depend on the radiation physics \citep{r96}.
In practice, we use simplified assumptions that capture the basic
behavior, and look for departures from the resulting predictions as guides
to improved modeling of the physics. For example, it is often assumed
that the gaps self-limit to a fixed potential drop \citep{a06} which means
that the $\gamma$-ray luminosity will be proportional to the open zone
current \citep{ha81}. One natural way to do this is to imagine radiation
reaction-limited particles passing through a charge-starved gap
spanning a fraction $w$ of the open zone of the pulsar magnetosphere.
In this paper we take the emission surface to be relatively thin, lying 
a fraction $w$ of the way from the last closed field line surface to 
the magnetic axis. For $w$ not too close to one, this approximates
the constant voltage limit if one takes the heuristic $\gamma$-ray luminosity 
$$
L_\gamma \approx (10^{33}{\rm erg/s} {\dot E}_{SD} )^{1/2} .
\eqno(1)
$$
This implies an efficiency $\eta=L_\gamma/{\dot E}_{SD}$ equal to 
the fractional gap width $w$. Realistic models must, of course, have a finite
thickness for the emission zone. We find that this produces relatively 
small perturbations to the light curve shape if $\delta w \la w$.

	The phase-averaged $\gamma$-ray flux for the Earth line-of-sight 
$F_{obs}$ is related to the total $\gamma$-ray luminosity by 
$$
L_\gamma = 4\pi f_\Omega F_{obs} D^2
\eqno(2)
$$
where $D$ is the distance to the pulsar \citep{wet09} and
$f_\Omega$ is the `flux correction factor'; an isotropic
emitter corresponds to $f_\Omega=1$. Notice that we need this quantity
to infer the actual pulsar $\gamma$-ray efficiency
$$
\eta= 4\pi f_\Omega F_{obs} D^2/{\dot E_{SD}},
\eqno(3)
$$
for comparison with model predictions. A prescription for $f_\Omega$
is given in Equation (4) of
\citet{wet09}; the solid angle corrections are applied correctly
in the values in that paper (albeit for the `Atlas'-style beam computation).

\subsection{Magnetic Field Structures}

    To start, one must choose a field line structure. In the earliest
light curve modeling \citep{mor83,cr94} static dipole magnetic
field lines were assumed. Emission was assumed beamed along these field lines,
which were taken to occupy the rotating frame, and the resulting 
aberrations and time delay formed the observed pulse. While this
``static'' (Stat) model had the right topology (closed zone plus 
flaring open zone),
and already predicted (for outer magnetosphere emission) the generic
double $\gamma$-ray pulse lagged from the low altitude radio pulsations,
it clearly is not a physically consistent picture.

\begin{figure}[h!]
\vskip 11.5truecm
\includegraphics{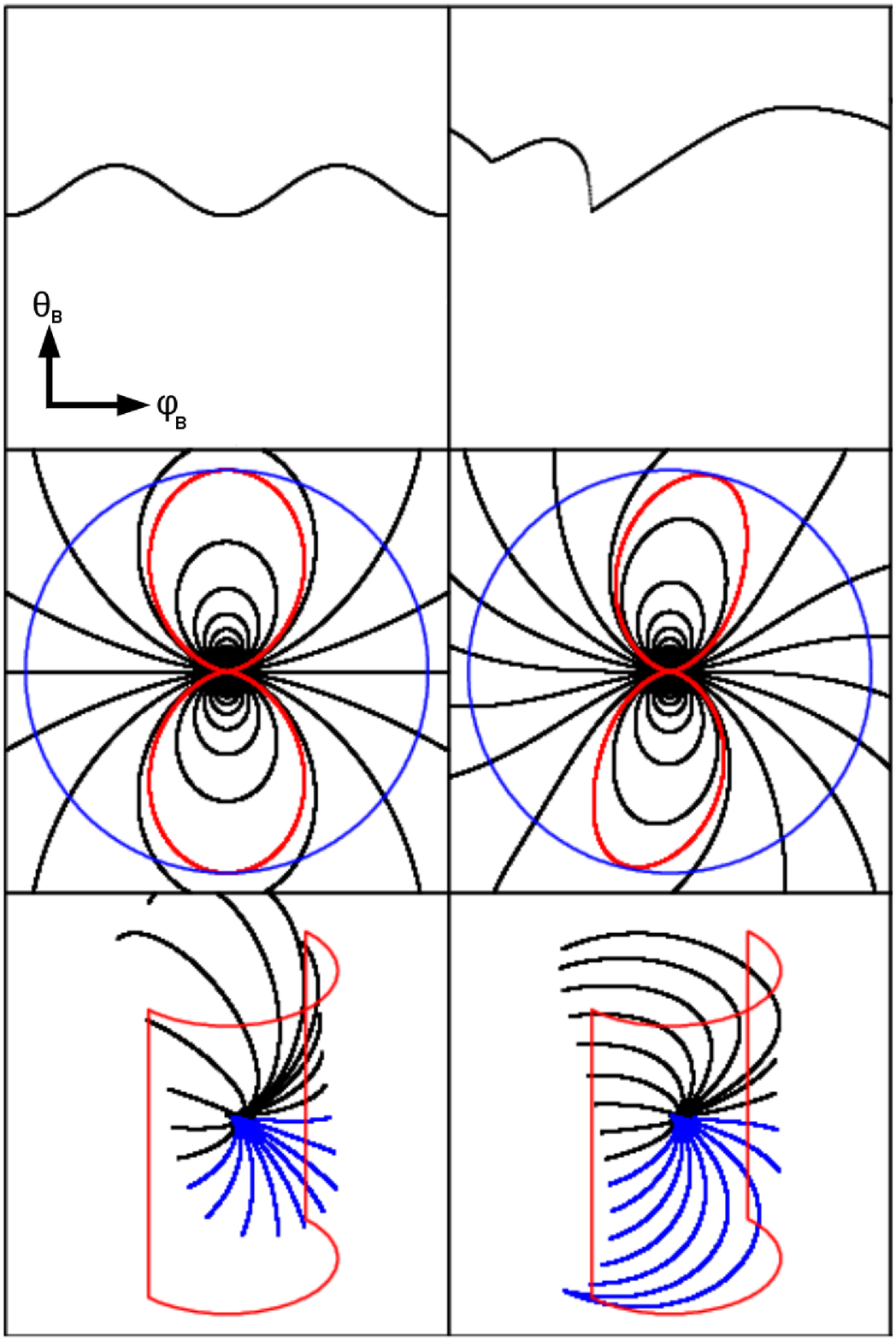}
\begin{center}
\caption{\label{Bgeom} Illustrations of the basic dipole field geometry. Left: Static,
Right: retarded point dipole (as used in the `Atl' and `PFF' computations). 
Top -- the position of the open zone boundary (polar cap boundary) in 
magnetic coordinates $\theta_B(\phi_B)$ for magnetic inclination $\alpha=70^\circ$.
Middle -- magnetic field lines in the equatorial plane for $\alpha=90^\circ$.
Bottom -- field lines for $\alpha=80^\circ$ above one polar cap with 
lighter (blue) showing the lines which cross $r_{NC}$, the `Outer Gap' lines.
}
\end{center}
\end{figure}

	The next step, introduced by \citet{ry95} and used widely
by other modelers, is to use the retarded dipole instead of the static
case. This is commonly referred to as the Deutsch (1955) field approximation.
However, at large radius from the star, this is effectively just a rotating 
point dipole which is most simply expressed by Kaburaki (1980) and
can be re-written as
\[
{\vec B} = -\left [ {\vec m}(t_r) + r {\dot{ \vec m}}(t_r) +r^2  {\ddot { \vec m}}(t_r) \right ]/r^3
\]
\[
\qquad + {\hat r} \left ({\hat r}\cdot \left [ 3 {\vec m}(t_r) + 3r {\dot{ \vec m}}(t_r) +r^2  
{\ddot { \vec m}}(t_r) \right ]/r^3 \right )
\]
\[
{\vec m}(t_r) = m ({\rm sin}\alpha\, {\rm cos} \omega t_r\, {\hat x} + 
{\rm sin}\alpha \,{\rm sin} \omega t_r \,{\hat y} +
{\rm cos} \alpha \,{\hat z})
\qquad (4)
\]
with $m$ the dipole magnetic moment evaluated at the retarded time $t_r=t-r/c$.
This is the field structure used in \citep{wet09} and so we refer to it
as the ``atlas'' (Atl) field. It has the virtue of having the intuitively 
expected `sweepback' as one approaches the light cylinder (Figure \ref{Bgeom}).

	This field is stationary in the non-inertial frame rotating with respect
to the lab frame, where particles are assumed to follow the field
lines. In \citet{wet09} and earlier computations by a number of groups, the 
particle velocity in the lab frame was incorrectly computed as a co-rotation induced
boost to a relativistic particle following the rotating field lines.
\citet{bs09a} correctly point out that the particle velocities in the two frames are instead
connected by a simple coordinate transformation. These authors assumed that the 
earlier computations (incorrectly) used the field (4) in the rotating frame. The
boost error of \citet{wet09} is conceptually different, but mathematically equivalent to
this assumption.  In practice, the coordinate transformation allows one to compute the 
velocity ${\beta^\prime_\parallel}$
along the magnetic field line for the particle forced to co-rotate with the
star at a highly relativistic lab frame velocity $|\beta_0|\longrightarrow 1$
$$
{\vec \beta_0} = {\beta^\prime_\parallel} {\hat B} + {\vec \Omega}\times {\vec r}/c
\eqno (5)
$$
which gives
$$
\beta^\prime_\parallel=-{\hat B} \cdot ({\vec \Omega}\times {\vec r}/c) +
\{ 
[{\hat B} \cdot ({\vec \Omega}\times {\vec r}/c)]^2 - ({\vec \Omega}\times {\vec r}/c)^2 +1
\}^{1/2}
\eqno (6)
$$
so that substituting into Equation (5) gives the direction of the particle 
as viewed in the lab frame. Note that $\beta^\prime_\parallel$ can become very small
for particles traveling nearly along ${\vec \Omega}\times {\vec r}$ for $r_\perp \sim
r_{LC}$, i.e. nearly tangent to the light cylinder in the `forward' direction.
In this case small effects of particle inertia or field line perturbations will
`break open' such field lines and the particles should leave the light cylinder
and contribute little to the pulse emission.  Although we do not follow these
physical effects, such field lines do not affect our light curves, since we taper
the emissivity (see \S 2.4 below). 

In any case, to have the plasma static in the rotating frame requires the
presence of the co-rotation charge density that produces the lab frame field
$$
{\vec E} = ({\vec \Omega } \times {\vec r} )\times {\vec B}/c.
\eqno(7)
$$
This is {\it assumed} present in the `Atlas' and similar computations,
and is similarly assumed here. We refer to the revised computation
as the `pseudo force-free' (PFF) case, 
since it contains co-rotation enforcing charges, but no currents.
These PFF computations,
for both TPC and OG-type models, give light curves slightly distorted from
in the `Atlas'-type computations and appear to match well to the sample 
antenna patterns and light curves of \citet{bs09a}. The principal effect,
as noted by these authors, is that for small $w < 0.02$ the low altitude
emission that dominates the second pulse of the TPC model has decreased
phase folding, making the Jacobian non-singular. There is a pulse peak
but without a true caustic there is no sharp pulse. This also weakens
somewhat the first TPC peak for such high ${\dot E}$ pulsars. For
pulsars with larger $w$ and for the higher altitude emission of the OG
model there are departures from the `Atlas'-style computation,
% call out the new appendix here....
but the effects are more subtle. In the Appendix we present arrays of 
example light curves for the TPC and OG models. These may be compared with,
and considered as updates to, the similar light curve panels in
Watters et al (2009). We also update the figures describing
the pulse width and flux correction factors and their variation with 
viewing geometry, for comparison with that paper.

% Time to comment on the primary fits file differences.

\subsection{Current-induced $B$ Perturbations}

	The gap-accelerated radiating charges alone produce some
currents in the open zone, and we expect the gap-closing pair
front to produce densities comparable to the co-rotation value for
at least some of the open field lines. This current will
perturb the magnetic field of Equations (4). 

	In the spirit of attempting analytic amendments to the 
vacuum model, we applied the approximate perturbation field computed
by \citet{mh09} for a pair-starved current flow in the open zone.
This is expressed in magnetic coordinates $(r_B,\theta_B, \phi_B)$,
where the perturbation amplitude $\epsilon^\prime$ determines
$B^\prime = \epsilon^\prime  \, [2 m\, \Omega/(r^2\,c)]$
with $m$ the magnitude of the star's dipole moment
\[
B^\prime_{r_B} = B^\prime \, s \, (1-s^2)^{-1}
\qquad\qquad
\]
\[
B^\prime_{\theta_B} = B^\prime 
[(1-s^2)^{-1} {{ \partial s}\over{\partial\theta_B}}+
\{{\rm sin}\theta_B(1-s^2)^{3/2}\}^{-1} {{ \partial s}\over{\partial\phi_B}}]
\]
\[
B^\prime_{\phi_B} = B^\prime 
[-(1-s^2)^{-3/2} {{ \partial s}\over{\partial\theta_B}}+
\{ {\rm sin}\theta_B (1-s^2)\}^{-1} {{ \partial s}\over{\partial\phi_B}}]
\quad (8)
\]
and $s={\rm cos}\,\theta_{rot} = {\rm cos}\,\alpha\,{\rm cos}\,\theta_B + {\rm sin}\,\alpha \,{\rm sin}\,\theta_B\,{\rm cos}\,\phi_B$.
This perturbation field is defined with respect to the {\it static} vacuum dipole.
We take $\epsilon^\prime$ positive for outward-flowing electrons 
(i.e. {\it negative} current) above the magnetic pole passing closest to
the Earth line-of-sight. This pole is generally inferred to produce the
low-altitude radio pulsar emission. If this pole has the opposite current
(i.e. opposite magnetic polarity) we have $\epsilon^\prime <0$. Of course
an electrodynamically self-consistent model will have a particular sign
of $\epsilon^\prime$ for, e.g. the positive magnetic pole, but our PFF model 
includes no current and so is insensitive to the sign of $B$; only
this perturbation current breaks the degeneracy.

\begin{figure}[h!]
\vskip 6.5truecm
\includegraphics{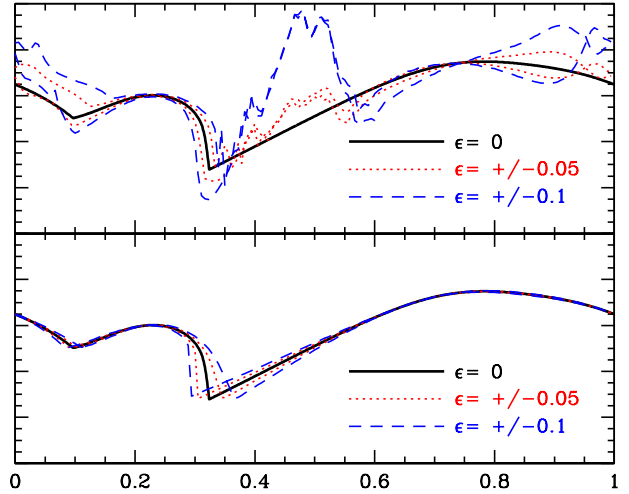}
\begin{center}
\caption{\label{footpts} Open zone boundary (polar cap) foot-points: magnetic colattitude
$\theta_B$ on the star surface as a function of magnetic azimuth $\phi_B$ for
the $\alpha=70^\circ$ PFF field with
current-induced perturbations. Top: the \citet{mh09} pair-starved field,
mapped to the retarded dipole.
Bottom: retarded dipole with perturbations from magnetic axis line currents.
Notice that the currents `twist' the field lines for both models,
shifting the `notch' at $\phi \approx 0.3$. Although the singularity 
in the \citet{mh09} field (upper panel) has been smoothed, we still 
see dramatic instability for lines directed near the rotation axis
($\phi_B \approx 0.4-0.6$ and  $\phi_B \approx  0.8-0.1 $).
}
\end{center}
\end{figure}

	To approximate the retarded solution, we matched this perturbation to our magnetosphere,
mapping $B(r_B,\,0,\,\phi_B)$ to the magnetic axis of the swept-back dipole. 
The perturbation field becomes singular along the rotation
axis where $s \longrightarrow \pm1$. By exponentially tapering this singularity
($\propto 1-e^{(|s|-1)/\sigma_s}$, with $\sigma_s =0.25$) we were able to integrate
the field lines to determine the last closed field line surface (see below). 
Unfortunately, for lines near $\phi_B = 0,\,\pi$ the residual effects of this
singularity still dominate, making the cap boundary at these azimuths unstable
(Figure \ref{footpts}). 

	Accordingly we proceeded with a simpler toy perturbation field
computed by integrating a constant line current passing through the star along 
the magnetic axis (the current extends to cylindrical radius 
$r_\perp=r{\rm sin}\theta=1.2R_{LC}=1.2c/\Omega$ to minimize end effects).  This perturbation field 
$$
\qquad\qquad{\vec B }^\prime({\vec r}) \propto \int_{B(r_b,0,\phi_B)} {\rm d}{\vec I} \times {\vec r}
\qquad\qquad\qquad\qquad (9)
$$
is treated as static in the rotating frame and is summed with the field of (4).
The effect of this simple perturbation field is dominated by the twisting
of field lines along the magnetic axis, as can be seen by the shift of the `notch'
foot-points of the last closed field line cap boundaries in Figure \ref{footpts}. This
rotation is also present in the \citet{mh09} field, but is
dominated by the singularity effects except for nearly aligned rotators 
$\alpha \approx 0$. For ease of comparison, we scale the perturbation amplitude
for fields (8) and (9) to the underlying dipole field (4) 
so that 
$$
\epsilon= \langle B'(0.5r_{LC},\theta,\phi) \rangle /
\langle | B(0.5r_{LC},\theta,\phi) | \rangle$$
where we average over a sphere at $r=0.5r_{LC}$. Again the sign of
$\epsilon$ is determined by the current; we expect opposite values for
the two poles.

\begin{figure}[h!]
\vskip 5.9truecm
\includegraphics{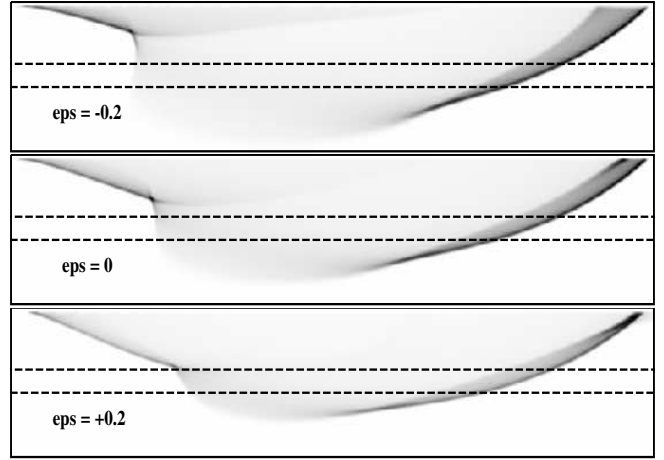}
%\begin{center}
\caption{\label{antpat} Antenna patterns for current-induced perturbations. 
Pulsar light curves are obtained from horizontal cuts across these images,
after normalization by $[{\rm sin}\zeta]^{-1}$. Here we show emission
from above the null charge surface (OG case) for $\alpha=60^\circ$,
$w=0.1$. The middle panel shows the unperturbed PFF structure. The upper
($\epsilon < 0$) and lower ($\epsilon >0$) panels show the effect of a perturbing line
current along the magnetic axis.  Note that the front caustic
is better closed for $\epsilon > 0$, but for negative $\epsilon$
the 'notch' structure is more pronounced. The dashed lines show
the range where the viewing angle $\zeta$ passes close to the low altitude
emission from the opposite magnetic pole, i.e. when
one expects a radio pulse. Measurement of light curve
perturbations can thus determine the sign of $B$ for a given magnetic
pole.
}
%\end{center}
\end{figure}

	None of these fields is fully realistic. In particular
the current-induced $B$ field perturbations used above are not complete
consistent field/current systems. Nevertheless, given that the vacuum
field geometry has already provided encouraging successes in 
approximating the observed light curves, it is useful to explore
how current systems can perturb the modeled beam patterns and pulse shapes.
As an example, Figure \ref{antpat} shows the 'antenna patterns' in the hemisphere 
containing emission from above the null charge surface (the OG case below).
The unperturbed beam pattern (middle) suffers opposite distortions in 
the two hemispheres.

%\vskip -0.5truecm

\subsection{Gap Locations}

	For each field structure, we assume the appropriate  
co-rotation charge density so that the field pattern is stationary in the
rotating frame.  We define the open zone by identifying
field lines tangent to the light cylinder (actually at slightly smaller radius
for numerical expediency). These field lines are traced back to the surface
where they define the open-/closed-zone boundary. In Figure \ref{Bgeom}  we show
the foot-point locations, equatorial plane field lines (for $\alpha=90^\circ$)
and the last closed field line structure for an inclined $\alpha=80^\circ$ rotator.
Our radiating surface is always in the open zone toward the
magnetic axis from this boundary. This plot shows the
static (`Stat') and point retarded dipole (`Atl' \& `PFF') field structures. 
The active field lines for the outer gap are lighter (plotted in blue).

With the emitting field lines defined, we must choose the extent of the
radiating surface.  For the TPC model, we
follow the original definition \citep{dr03,dhr04}, taking the
emission surface to be the full set of last-closed field lines,
but stopping emission once the radial distance is
$r> r_{LC}$ or once the distance from the rotation axis exceeds
$r_\perp > 0.75 r_{LC}$. This cut-off is required
to avoid the higher altitude pulse components (as used in the OG
picture), otherwise the modeled light curves generally 
have three or four pulse components.  As in \citet{wet09}
we make this model more physical by placing the emission surface
in the open zone, separated from the last closed surface by the
thickness $w$ appropriate to the pulsar ${\dot E}$ (Equation 1).

	In the outer gap (OG) model radiation is produced on
field lines crossing the null-charge surface (${\vec \Omega} \cdot 
{\vec B}(R_{NC})  = 0)$, with emission starting near this crossing point
and extending toward the light cylinder. Again, we assume
a gap thickness $w$.  Field lines that do not cross the null charge
surface before $r_{LC}$ do not radiate.

	The outer gap emission should be dominated by a section
of the flared cone of field lines above a given pole. For these simple
vacuum-based models, some field lines just inside the last closed surface
extend for a very long distance before exiting the light cylinder
or crossing the null charge surface.  For example, some field lines
arc over the star and cross ${\vec \Omega} \cdot {\vec B}=0$ for the first time
far from the star with a path length (in units of $R_{\rm LC}$) $s \gg 1$.
These field lines define a disjoint, `high altitude' second gap surface, that does
not in most cases appear to be active (although the HF pulse 
components of the Crab might represent such emission; Moffet \& Hankins 1996). 
Physically, we expect that such lines arcing near $R_{LC}$
will be opened in a real magnetosphere with currents and particle 
inertia. Also, gaps on field lines starting at large $r_{NC}$ will likely
be inactive if the soft emission needed to close the gap arises
from other field lines at much smaller $r_{NC}$. Accordingly, as in \citet{ry95},
we apply a path-length $s$ cutoff in the gap surface emissivity.
Here the functional form used (for $s>1+2s_{\rm NC,min}$) is
$$
F(s) \propto e^{-[(s-2s_{\rm NC,min}-1)/\sigma_s]^2}, \qquad \sigma_s=0.1
\eqno (10)
$$
with $s_{\rm NC,min}$ the lowest null charge crossing for any
active field line and all
distances in units of $R_{LC}$. A detailed surface emissivity 
weighting awaits a physically realistic spectral emission model.

%Note some general patterns of the PFF models vs. the atlas models.

	While this work was being prepared for publication, a new discussion
of light curve formation in fully force-free models has been presented
by Bai \& Spitkovsky (2009b). These authors note the limitations with pure
vacuum modeling mentioned above. They also argue that in fully force-free
magnetospheres if the radiation is associated with the return current sheet,
neither TPC nor OG pictures produce realistic light curves, and propose
a new `Annular Gap' (AG) geometry that has emission extending beyond the
light cylinder. The foot-points of the AG field lines are the same as
those used here for the TPC geometry (for a given $w$) but in 
\citet{bs09b} the co-rotating pulsed emission is taken to extend well beyond the
the light cylinder.  Thus, this model covers our OG zone
plus additional emission at both high and low altitudes. As we see below,
the vacuum dipole field (possibly with current-induced B perturbations)
does a rather good job of matching the observed pulsations -- it
will be interesting to see if the force-free numerical fields  using
the AG or OG geometries can match the data comparably well.

\subsection{Computational Grid}

	We identify the last closed field line surface (those field lines
tangential to the speed of light cylinder) and trace these
field lines to the polar cap surface at $r_\ast$.  We follow the field line
structure to $< 10^{-4} r_{LC}$, allowing good mapping even for pulsar 
periods of $\ge 1$s.
A single magnetosphere model can serve a range of pulsar periods, since
the OG and TPC model features are computed as distances in fractions
of $r_{LC}$. The pulsar period only affects $R_\ast/R_{LC}$ and hence
the inner boundary of the emission zone.

Models are computed for all magnetic inclinations $\alpha$.
Photons are emitted along the particle path in the lab
frame (moving tangent to the local field line in the co-rotating frame)
and then phased, including light travel time across the magnetosphere,
to add to the skymap antenna pattern, as would be seen by distant
observers. We assume uniform emissivity along all active field lines
(except for the taper at large pathlength $s$
as noted above). In practice we trace the trajectories for $\sim 10^4$
photons per field line, which produces adequately smooth light curves for the
present computation. Models are computed for a range of gap widths from 
$w=0.01$ to $w=0.3$, to accommodate a range of pulsar ${\dot E}$.

\subsection{Lightcurve Fitting}

	Our next task is to take this family of light curves and compare with
the observed pulse profiles. We take our $\gamma$-ray data from the full band
$E>0.1$\,GeV pulse profiles published in \citet{psrcat}.
These use the first six months of {\it Fermi} data and
give good statistics for the relatively bright pulsars considered here.
In this {\it Fermi} pulsar catalog, a background level is estimated
from the surrounding sky. Interestingly, some pulsars have significant
steady (DC) emission. Matching this emission is an important test of the
model -- accordingly we fit to the total emission, pulsed and un-pulsed,
after subtracting the background, as estimated in \citet{psrcat}. In particular,
we do {\it not} fit the un-pulsed flux as a separate degree of freedom --
the excess above the LAT pulsar catalog background is determined by 
the magnetosphere model.

	We take the measured pulsar ${\dot E}$ and compute $w$ according
to Equation (1), then use the closest value from the model grid.
For this grid we take all $\alpha$ (in one degree intervals) and
fit the model light curve for each $\alpha$ and $\zeta$.  To compare with
the data we bin a given model curve into the phase bins (25 or 50/cycle) 
defining the light curve for the pulsar in \citet{psrcat}. The total model
counts are normalized to the background-subtracted data counts 
and we then search for the best-fit model to the 
combined (pulsar plus background emission) light curve.

\begin{figure}[h]
\vskip 6.1truecm
\includegraphics{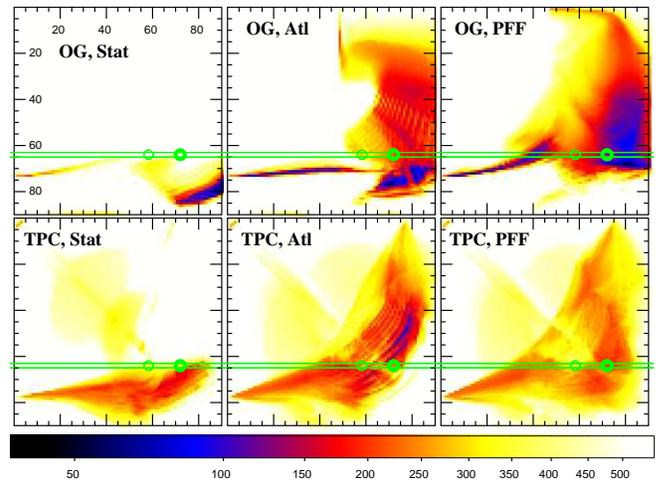}
\caption{\label{velacomp} Goodness of fit $\chi(\alpha, \zeta)$ surface 
for Vela with several models. Low $\chi$ (dark) represent the best
fits.  The three columns give results (Left to Right) for Stat, Atl and PFF field 
structures. Top are OG models, bottom are TPC.  The green lines show 
$\tilde\zeta$ allowed by {\it CXO}\, PWN fitting. The ellipses give the combined 
constraints including radio polarization data.}
\end{figure}

	A variety of fitting statistics work well. Simple $\chi^2$ fitting
generally finds the best solutions, but is controlled by the more numerous
off-pulse and `bridge' emission phase bins and so occasionally provides poor discrimination
against models with unacceptably weak peaks. Choosing the model with
the smallest maximum model-data difference for any bin ${\rm Max[Abs({\rm Mod_i-D_i})]}$
does an excellent job of matching the peaks, but less well on the bridge flux.
We find that an exponentially tapered weighting of the largest model-data
differences is robust and produces sensible results. Thus we minimize
$$
\chi_n = {{({\rm Mod_i-D_i})^2} \over D_i} \, {\rm e}^{-i(|{\rm Mod_i-D_i}|)/n}
$$
with $i(|{\rm Mod_i-D_i}|)$ the index of the phase bin differences, sorted large 
to small. Small $n \ll 1$
just uses the biggest difference bin (and thus focuses on the peaks), large $n$ goes 
over to classical $\chi^2$ weighting.
The method is only weakly sensitive to $n$ and an e-folding weighting $n\approx 3$
worked well.

\begin{figure}[t!!]
%\vskip 8.0truecm
\vskip 4.8truecm
\includegraphics{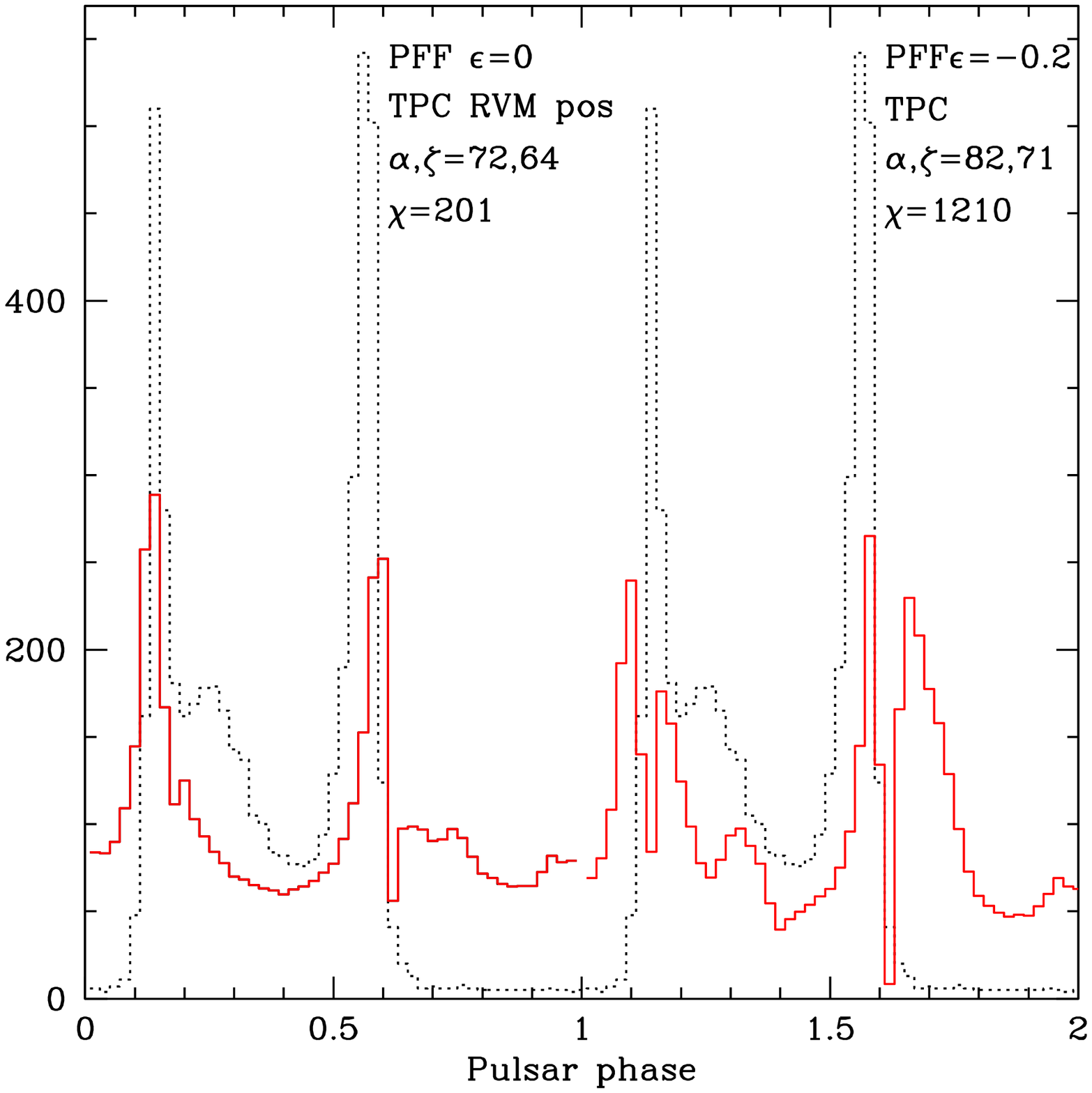}
\includegraphics{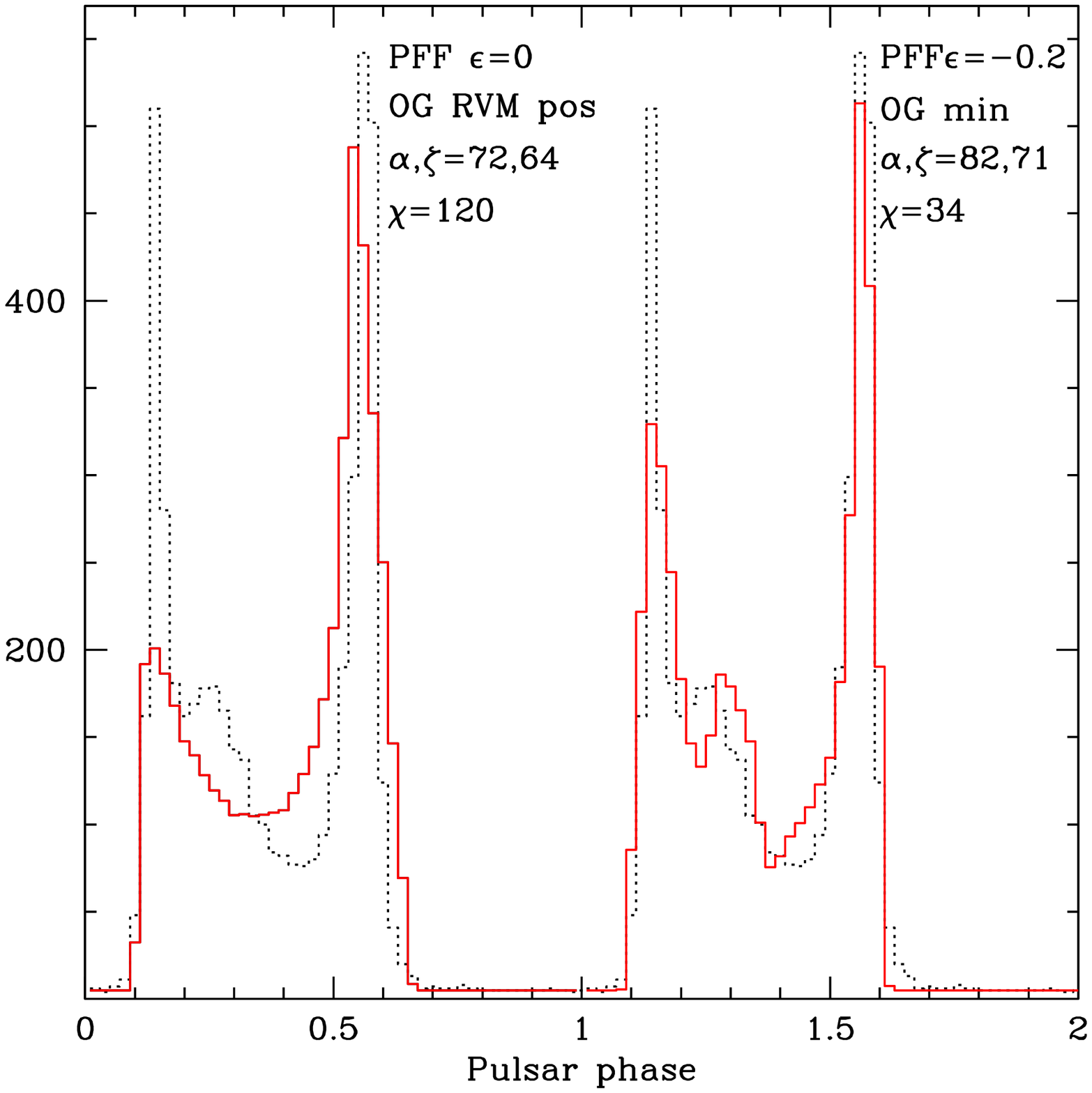}
\caption{\label{velalc} Pulse profiles for the TPC (left) and
OG (right) models for Vela using the PFF field structure.  The dotted pulse
profiles show two periods of the LAT ($E>0.1$GeV) Vela data from
\citet{PSRcat}. The first model 
pulse in each panel is for the PWN/RVM-fit $\alpha$, $\zeta$, with no current perturbation
while the second is for the parameters of the global OG fit minimum. The TPC 
fit quality varies only weakly with angle and does not improve substantially
away from the RVM values.}
\end{figure}
\medskip

	Our reference phase $\phi=0$ occurs when the dipole axis,
spin axis and Earth line of sight all in the same plane.  It is generally assumed that
radio emission arises at relatively low altitudes (but see Karastergiou and Johnston 2007),
and that the radio pulse occupies a fraction of the open zone
field lines at low altitudes. In \citet{psrcat} phase zero is taken
to be the amplitude peak of the main radio pulse. However, the radio
pulse profile often exhibits `patchy' illumination of the emission zone \citep{lm88}.  
Thus the 
radio peak may not mark the phase of the closest approach of the magnetic axis.
Additional geometrical information comes from the linear polarization sweep
of the radio pulse, with the maximum sweep rate associated with the
line-of-sight passage past the magnetic axis. This maximum often occurs
well away from the intensity peak. Finally, if the radio emission
arises well above the star surface, relativistic effects can
perturb the phases of the closest approach (ideally pulse intensity maximum)
and sweep rate. \citet{bcw91} argue that the approximate phase shifts
are 
$$
\phi_{\rm Int} \approx -r/r_{LC} \qquad\qquad \phi_{\rm Pol} \approx +3r/r_{LC} .
$$
Thus polarization information can give additional constraints on the
absolute phase, but altitude (plus mode changing and other sweep
perturbations) can lead to substantial additional uncertainty.
Accordingly, we have the option to fit light curves with the model phase free.
Generally, we allow $\phi$ to shift by as much as $\pm$0.1 of a rotation, and retain
the best fit model. When no radio pulse is known we allow full
$\delta\phi=\pm 0.5$ freedom in the pulse phase. The phase for the best-fit
model is recorded in the ($\alpha$, $\zeta$) map, as well as the
model light curve and properties of the pulsed emission, for use in
further analysis. Recalling that the radio phase $\phi=0$ in \citet{psrcat}
is set at the pulse intensity peak, we expect that our fit, which
estimates $\phi_0$ for the true magnetic axis, will tend to give
small $\delta \phi > 0$ for emission from finite pulse heights. However,
we expect that $\delta \phi$ will be less than $\phi_{\rm Pol}$.
In practice, patchy pulses and polarization sweep uncertainties make
$\delta \phi$ of any modest amplitude plausible.

	In considering the values of the fit statistic $\chi_n$, we must
remind the reader that these are not complete models.  Detailed
radiation physics will certainly
change the light curve shape from the uniform weighting assumed here.
Also, the lack of self-consistent magnetospheric currents must, at some level,
cause light curve shape errors. This, coupled with the excellent
signal-to-noise of the {\it Fermi} light curves {\it guarantees} that
these will not be statistically ``good'' fits. Monte Carlo experiments with
the $\chi_n$ fit statistic show that for Poisson realizations of a 50 bin 
light curve, the typical value is $\langle \chi_3 \rangle = 7.5$ while 99\% and
99.9\% values are 15.5 and 19. Thus fit differences of $\Delta \chi_3 \approx
8$ are {\it statistically} significant at the normal $2.5 \sigma$ level.
This gives a guide to interpreting the relative goodness of fits to the various
models.

\section{Fit Examples}

	We start with the archetype $\gamma$-ray pulsar, Vela. For this 
pulsar we show the fit statistic $\chi$ surfaces for the three approximate dipole
field structures (Stat, Atl, PFF) discussed above, with darker colors 
indicating better model fits. The upper row is for OG models, the lower 
row for TPC. 
For OG all field structures show a similar topology, with best-fit solutions
at $\zeta \approx 70-80^\circ$ and $\alpha \approx 60-90^\circ$, and no
solutions at small $\alpha, \, \zeta$. The swept-back models (Atl and PFF)
have a tail of solutions to smaller $\alpha$. The TPC scenario can produce
models for nearly all $\alpha, \, \zeta$. However, except for the `Atlas'
geometry which, as also noted by \citet{bs09a}, has unrealistically sharp
caustic pulses for small gap width $w$, the models produce much poorer
fits than the OG scenario, as they have weak pulses and far too much
off-pulse (DC) emission.

\begin{figure*}[t!!]
\vskip 5.truecm
%\special{psfile=f5.eps hoffset=-25 voffset=-130 vscale=50 hscale=50}
\includegraphics{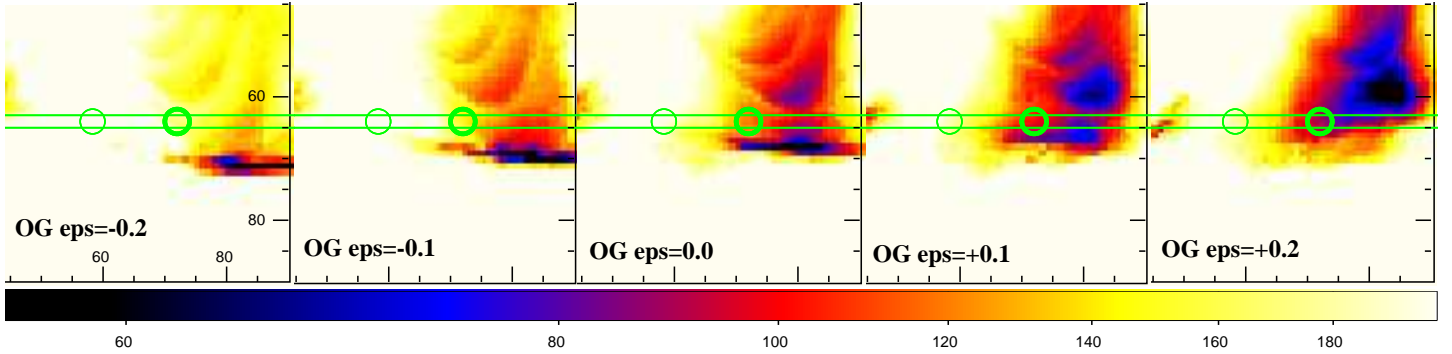}
\caption{\label{velaB}  Comparison of the lower right quadrant of the
OG fit planes for models with 
magnetic axis current-induced $B$-field perturbations. Left to
right: $\epsilon=-0.2,-0.1,0.0,+0.1,+0.2$. The plots
show the range $45^\circ \le \alpha \le 90^\circ$, $45^\circ \le \zeta \le 90^\circ$.}
\end{figure*}

\subsection{External Angle Constraints}

	The model fits in ($\alpha, \, \zeta$) space are especially useful
when we have external constraints on these angles. Some care is needed, 
however, in applying angle constraints from other wavebands.  For Vela and a number
of other young pulsars with bright X-ray pulsar wind nebulae (PWNe) we have
been able to place relatively model-independent constraints on viewing angle
by fitting for the inclination $\tilde{\zeta}$ of the relativistic Doppler-boosted
PWN torus \citep{nr08}. This method does not, however, distinguish the sign
of the spin axis. Therefore the X-ray torus fits allow two possible
viewing angles between the positive spin axis and the Earth line-of-sight
$\zeta=\tilde{\zeta}$ and $\zeta^\prime=180^\circ-\tilde{\zeta}$. 
In fact, in \citet{nr08}
the torus position angle was restricted to be $0^\circ \le \Psi < 180^\circ$,
so the angles quoted in this paper may represent either $\zeta$ or
$\zeta^\prime$.

	In the context of the rotating vector model \citep{rc69}
radio polarization data can also constrain the viewing angles.
In most cases, the small range of phase illuminated by the radio
pulse allows only an estimate of the magnetic impact parameter
$$
\qquad\quad\beta = \zeta - \alpha \approx 
{\rm sin^{-1}} [{\rm sin} \alpha/({\rm d}\Psi/{\rm d}\phi)_{\rm max}]
\qquad\qquad (11)
$$
where the maximum rate of the polarization position angle sweep $\Psi(\phi)$
occurs at $\phi_{pol}=0$, near the closest approach to the magnetic axis. Here the sign
of the sweep is meaningful, determining whether the line of sight is closer to 
or farther from the positive rotation axis than the observed magnetic pole (at 
inclination $\alpha$). Thus, when we have an independent PWN estimate of
$\tilde{\zeta}$, we have two possible $\alpha$ values -- e.g. for a negative sweep
we get $\beta < 0$ and $\alpha=\zeta-\beta$ or $\alpha=\zeta^\prime - \beta$.
In this paper, we tabulate the possible values of $\zeta$ and $\zeta^\prime$, but 
save space by plotting only in the positive rotation hemisphere. Thus there
are two values possible for $\alpha$ from Equation (10), but when $\alpha$
or $\zeta > 90^\circ$, we reflect back into the positive hemisphere for plotting
purposes.

	Occasionally, when the radio pulse is very broad or when the pulse profile
presents an interpulse, the radio polarization can make meaningful estimates
of both $\alpha$ and $\zeta$, from fits to the full polarization sweep
$$
{\rm tan}(\Psi + \Psi_0) = 
{{{\rm sin}\alpha\, {\rm sin} (\phi-\phi_0)} \over 
{{\rm sin}\zeta\,{\rm cos}\alpha - 
{\rm cos}\zeta \,{\rm sin}\alpha\,{\rm cos}(\phi-\phi_0)}
}
$$
where the polarization has the absolute position angle $\Psi_0$ at $\phi_{Pol}=\phi_0$.
\citet{keith09} have recently presented several examples. When the radio
illumination is insufficient for a full solution, we can still break the
degeneracy of the fits using the X-ray $\tilde{\zeta}$ constraints.
Finally, if the magnetic inclinations orientations are, at least initially, isotropic
we expect the inclination choice giving $\alpha$ closest to $90^\circ$ to be preferred
on statistical grounds.

In Figure \ref{velacomp} the two possible RVM $\alpha$ solutions for Vela
\citep{jet06}
consistent with the PWN data are shown by the green circles. Note that the
OG and TPC models have minima close to the solution at larger $\alpha$ 
(bold circle). In general, with constrained $\alpha$, $\zeta$  values
one can make a discrimination between the models. Note that for the OG model
the position of the minimum is closest to the preferred angle for the 
relatively physical `PFF' case. The values of the fit statistic are substantially
worse for TPC at all locations near these preferred angles (Table 1).
Figure \ref{velalc} shows the PFF light curves at the RVM angles, and at 
the global fit minimum, for the TPC and OG scenarios.

\subsection{Testing Field Perturbations}

	The best-fit models in Figure \ref{velacomp} are several $\sigma$ off of the PWN/RVM
angles. This suggests either that there are systematic errors in these angle estimates or
that the true magnetosphere structure is slightly different than the PFF estimate
above. The measurement of such offsets for a number of pulsars should enable us
to map the required geometry differences and test physical models for their origin.
As an example, here we illustrate the perturbing effect of magnetospheric currents.
Figure \ref{velaB} shows the $\chi$ plane for fits including a magnetic
axis current (Equation 9). 
The current shifts the fit minimum and distorts the light curves.
For a best fit in the RVM-determined region positive currents are required for
the OG model. The best models within $\sim 1 \sigma$ of the PWN+RVM angles also
prefer a small positive current. For negative $\epsilon$ the best fit region
shifts further from the externally determined angles, although the match to
the pulse shape is quite good, reaching a global minimum ($\chi=34$)
at $(\alpha, \, \zeta) = (82^\circ, 71^\circ)$ for $\epsilon=-0.2$.
Since the global minimum does not shift precisely to the PWN/RVM
angles, we infer that (unsurprisingly) the true field geometry is only 
approximated by this simple model.
The TPC model fits are not improved by including finite $\epsilon$;
the best minimum anywhere near the external PWN/RVM angles has 
$\chi \approx 200$.

To summarize, for Vela with the external constraint, PFF is clearly preferred
and the best fit models are relatively close to the known $\alpha$, $\zeta$.
The rotation axis is inferred to be oriented out of the plane of the sky,
we are viewing at $\zeta \approx 65^\circ$  and the magnetic inclination
is large at $\alpha \approx 75^\circ$. Positive-current field perturbations
are preferred. The $\gamma$-ray pulse fits imply that $\phi_0$ lies earlier
than the radio peak (and earlier than the maximum sweep rate). This would
suggest that Vela has patchy radio emission dominated by the `trailing' side of
the radio zone. 

\begin{figure}[h!!]
\vskip 6.1truecm
\includegraphics{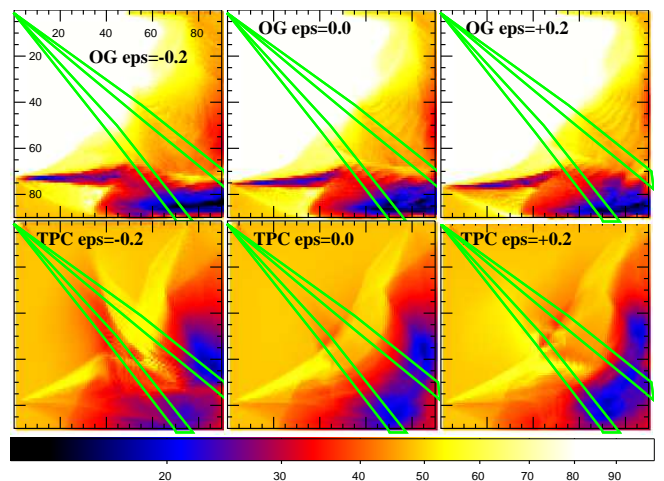}
\caption{\label{1952comp} Goodness of fit surface for PFF models
of PSR J1952+3252,
Left to right: $\epsilon=-0.2,\, 0.0,\, +0.2$, top: OG, bottom: TPC. The
regions indicated by the maximum polarization sweep rate 
${\rm d}\Psi/{\rm d}\phi|_{\rm max}$ are shown as two green wedges.}
\end{figure}

\subsection{Other EGRET Pulsars}

	The next most energetic pulsar in our test set is PSR J1952+3253.
As for Vela, the three field structures give quite similar $\chi$-plane
maps. In (Figure \ref{1952comp}) we show the $\chi$ maps for the PFF model
with three values of $B^\prime$.  This pulsar has
two narrow peaks separated by $\Delta=0.49$ in phase, so we find that
the TPC model can produce reasonable solutions, as indicated by
the relatively dark regions in the lower panels.
This object is presently interacting with the shell of its supernova
remnant, CTB80, and so produces a PWN bow shock rather than a torus,
precluding a $\tilde{\zeta}$ measurement. However from the max sweep rate
of the radio polarization from Weisberg et al. (1999), we can infer
a constraint on the magnetic impact parameter (two green diagonal bands). 
The sweep maximum appears to lie towards the tail of the broad radio
pulse, consistent with our fit $\delta\phi \approx 0.05$; again the radio
pulse peak would represent patchy emission, in this case from the leading edge
of the emission region at relatively low altitudes.

	Both models prefer relatively large $\alpha$ and $\zeta$,
and each can produce acceptable light curves within the region allowed 
by the radio data.  If improved radio measurements or other angle
constraints can be obtained, it may be possible to select between these
two models for this object and also measure the altitude of the radio
emission zone.
%This emphasizes that the best chance of differentiating pulsar
%magnetospheric structures occurs when estimates of both $\alpha$
%and $\zeta$ are available.

\begin{figure}[h!!]
\vskip 8.9truecm
\includegraphics{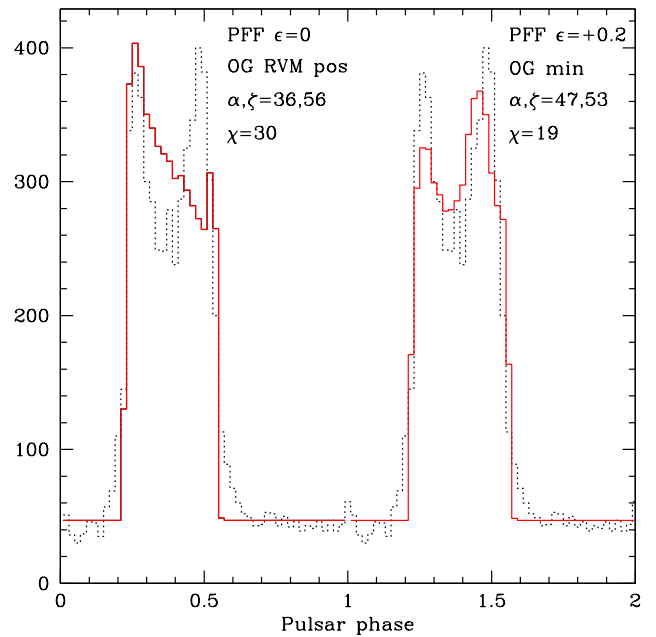}
\caption{\label{1706lc} Light curves for two PFF models for PSR J1709$-$4429.
The model at the RVM position has a rather weak second peak and requires
a relatively large $\delta \phi$. A model at the PWN $\zeta$ but
slightly larger $\alpha$ provides the best fit at $\delta \phi\sim 0.$
}
\end{figure}

	Next and only slightly less energetic is PSR B1706$-$44 = PSR J1709$-$4429.
This object, in contrast to PSR J1952+3252, has a narrow double pulse with
$\Delta=0.25$. Here OG models clearly prefer the small $\alpha$ RVM solution.
While the best OG fits have $\chi \le 30$, the TPC models in the
PWN-determined $\zeta$ strip are poor with $\chi > 90$. Polarization
sweep data are limited for this pulsar. The published plots in
\citet{jet06} give $\alpha \approx
34^\circ$, which is best matched with negative currents $\epsilon \le 0$.
These fits however, imply a surprisingly large offset for the magnetic
pole of $\delta \phi = -0.04= 14^\circ$, which would put the magnetic axis
at the leading edge of the radio pulse. Intriguingly, the best fit of all
parameters is found with a OG PFF $\epsilon=0.05$ model at $\zeta=53^\circ$,
$\alpha=47^\circ$ ($\chi=19$). This very good light curve (Figure \ref{1706lc})
has the
magnetic axis near the radio peak with $\delta \phi =-0.01$, but would have
a polarization sweep rate $\sim 3\times$ larger that that seen in present
radio data. Again, improved radio polarization data might help shed light 
on the preferred inclination angle.

\begin{figure}[h!!]
\vskip 6.0truecm
\includegraphics{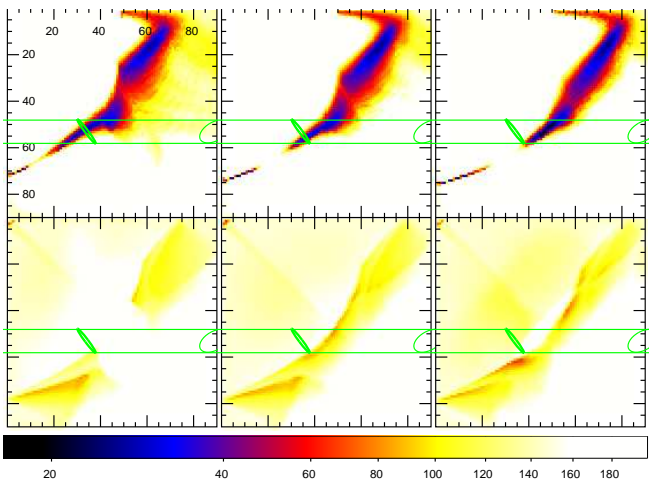}
\caption{\label{1706comp} Goodness of fit surface for PSR J1709$-$4429,
panels as for Figure \ref{1952comp}.
The X-ray determined $\tilde\zeta$ is shown by the horizontal band. For
this pulsar the smaller $\alpha$ RVM solution is clearly preferred.
}
\end{figure}

	The next pulsar, PSR B1055$-$52 = PSR J1057$-$5226, is appreciably older.
It shows a pulse that is nearly a square wave in the full energy
band, although the energy-dependent light curves of \citep{psrcat} indicate
that it is likely a tight double of separation $\Delta\approx 0.2$
with strong bridge emission. The pulsar is too old to show a bright PWN,
so no $\zeta$ is available from the X-rays. Luckily in the radio it has 
both a very broad pulse and interpulse and radio polarization fitting
can strongly constrain both $\alpha$ and $\zeta$. We adopt the values
from \citep{ww09} and note the phase of the maximum polarization sweep
is clearly at $\phi=0.08$ (later) than plotted in \citep{psrcat}. Indeed,
we find that the model fits prefer positive $\delta\phi$, although the
value is slightly larger than expected.
The peaks on the model light curves are also stronger
than in the data for these parameters; the match can be significantly
improved for models with finite $\delta w$ (not detailed here). As noted by \citet{wet09},
no viable TPC models lie anywhere near the RVM-determined angles. 
For OG there is a shallow fit minimum centered on the RVM angles.
Superposed on this are $\chi$ stripes, caused by the relatively
coarse (25 bin) LAT light curve. With improved LAT data the
fits should become more discriminating.

For the TPC model there is a large region with modest $\chi$ values 
at small $\alpha$ and $\zeta$, with the best solution listed in Table 1. 
However these models are not only inconsistent with the radio
data, but all produce single pulses with one broad,
approximately Gaussian, component
and are unlikely to represent the true solution. If one allows any
viewing angle, exceptionally good fits can in fact be found for OG with
$\chi$ slightly below the statistical minimum value. Despite an
excellent match to the observed $\gamma$-ray pulse, this
is also unlikely to represent the true solution. This underlines
the fact that sensible solutions must be compatible with
external angle constraints and, when these constraints are applied with care,
one may use the multi-wavelength data to rule out otherwise viable pulse models.

\newpage
\begin{figure}[t!!]
\vskip 6.5truecm
\includegraphics{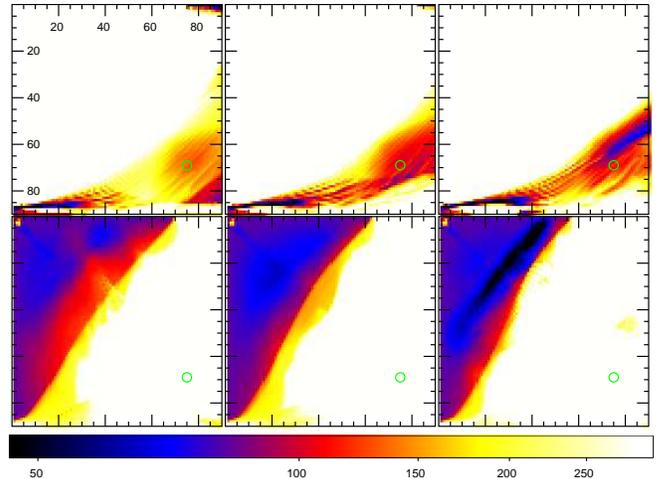}
%\special{psfile=f6_1055_lc.eps hoffset=333 voffset=-57 vscale=33 hscale=33}
\caption{\label{1055comp} Goodness of fit surface for PSR J1057$-$5226,
panels as for Figure \ref{1952comp}.
For this pulsar, interpulse emission allows a good RVM fit for both
the inclination angle $\alpha$ and viewing angle $\zeta$ (green circle).}
\end{figure}

	Finally, we discuss Geminga, the archetype of the old $\gamma$-selected
pulsars, and the only one known prior to the launch of {\it Fermi}. This
object does show an X-ray PWN, but like many old objects it is dominated by
bow shock structure \citep{psz06} and so it will be very difficult to
extract torus/jet constraints on $\zeta$. However we do know that
despite very sensitive radio searches, no convincing detection of pulsed
radio emission has been found, implying that $|\beta|=|\zeta-\alpha|$ is 
large. It has also been suggested that the thermal surface pulsations
are most consistent with a nearly aligned rotator 
(small $\alpha$, Pavlov, Sanwal \& Zavlin 2006).
%Oops probably should reference Caraveo and co here!!!

\begin{figure}[b!]
\vskip 6.1truecm
\includegraphics{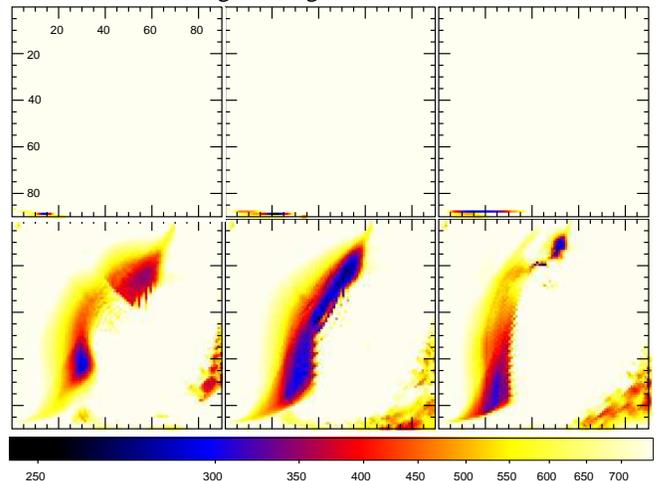}
%\special{psfile=f7_0633_lc.eps hoffset=333 voffset=-57 vscale=33 hscale=33}
\caption{\label{0633comp} Goodness of fit surfaces for PSR J0633+1746 (Geminga),
panels as for Figure \ref{1952comp}.
No quantitative external angle constraints are available, although
the absence of radio emission suggests that $\beta=\zeta-\alpha > 15^\circ$.
TPC shows a number of acceptable solutions; OG has few (albeit better)
solutions at small $\alpha$, large $\zeta$.
}
\end{figure}

\begin{figure}[b!!]
\vskip 8.9truecm
\includegraphics{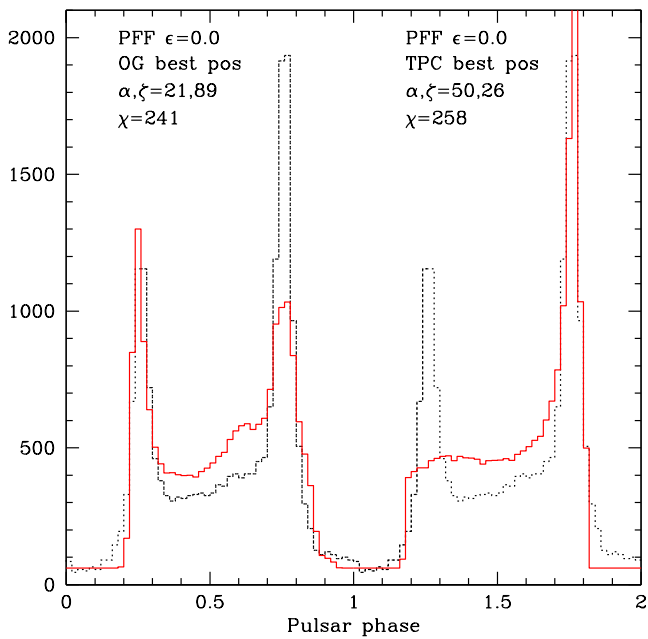}
\caption{\label{0633lc} Light curves for two PFF models for PSR J0633+1746
(Geminga). The left model is for the best OG solution, the right model
for the best TPC solution away from $\beta \approx 0$.
}
\end{figure}

	Here, with no information from the radio phase we must allow the 
$\phi$ to be freely fit. The results are interesting. For all 
$B$ field structures, and in particular the PFF structure, the OG model only
fits satisfactorily in a small region at small $\alpha$, large $\zeta$. This
means that one is viewing a nearly aligned rotator from near the spin
equator, i.e. very far from the magnetic axis. This is in good
agreement with the expectations of \citep{ry95} and guarantees that the
pulsar should be radio faint. In figure \ref{0633comp} we show surfaces computed
for $w=0.18$ and three trial $\epsilon$ values.  TPC shows best solutions over a range 
of angles and $\epsilon$. However, these have $\Delta \chi > 17$ larger
than the best OG solution and show only a single peak
with bridge emission (Figure \ref{0633lc}).  Thus while only few OG models are suitable, the 
light curve shape fits better and the viewing angles for Geminga are very 
well constrained.  It would be very interesting to measure $\zeta$ and/or $\alpha$ from
data at other wavelengths.

	In Table 1, we collect these model fit results, giving the values for
the externally constrained fits for the unperturbed $\epsilon=0$ field 
geometry for both
models. In addition we quote best fits close to the RVM angles and the best
global fits, when appropriate. For Geminga, with no strong radio constraint
we only list the global best fits for each model.

\begin{figure}[h!!]
\vskip 8.0truecm
\includegraphics{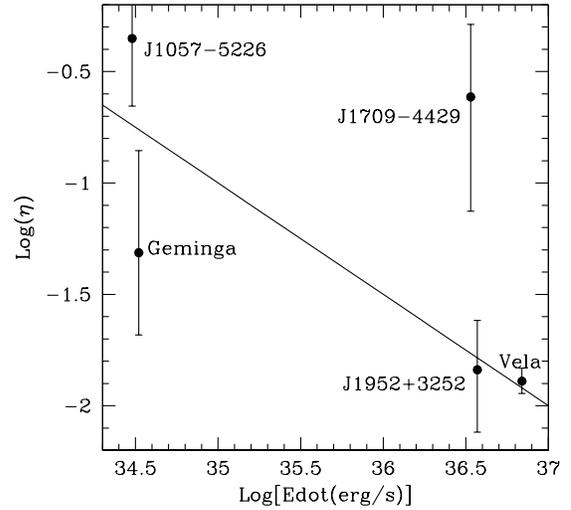}
\caption{\label{eff}Pulsar $\gamma$-ray luminosity, corrected from the observed
phase-averaged flux to all-sky vs. the spindown luminosity. OG/PFF/$\epsilon=0$ points.
%$f_\Omega=1$ open circles.
}
\end{figure}

\begin{deluxetable*}{lccccclrclrrcc}[b!!]
\tablecaption{\label{PulsarParameters} PFF Model Angles and Efficiencies }
\tablehead{
  \colhead{Name} & \colhead{log(${\dot E}$)} & \colhead{$w$} & \colhead{$d\tablenotemark{a}$}
    & \colhead{$\alpha/\zeta$} & \colhead{$\alpha/\zeta^\prime$}
    & \colhead{$\alpha/ \zeta/\epsilon$} & \colhead{$\chi$/$\delta\phi$} & \colhead{$f_{\Omega}$}
    & \colhead{$\alpha/ \zeta/\epsilon$} & \colhead{$\chi$/$\delta\phi$} & \colhead{$f_{\Omega}$} 
    & \colhead{$F_{>0.1{\rm GeV}}$\tablenotemark{a}} & \colhead{$\eta_{OG}$} \cr
%   & ${\rm erg\,s^{-1}}$ & & & Ext &  & TPC & TPC & TPC & OG & OG & OG &  &
   & ${\rm erg\,s^{-1}}$ & & &\multicolumn{2}{c}{$\longleftarrow$ External $\longrightarrow$\qquad}&\multicolumn{3}{c}{$\longleftarrow$\qquad TPC\qquad $\longrightarrow$ } & \multicolumn{3}{c}{$\longleftarrow$\qquad OG\qquad $\longrightarrow$} & &
}
\startdata
%    ALERT the dleta phi value here are OPPOSITE those reported by the fit program and plotted in the LC fit plots.
%Crab         & 38.7 & 63    &     & 0.40  & 0.111 & 60-90 & 63 & 1.0-1.25 & 75-90 & 63 & 1.0-1.25\\
Vela         &36.84 &0.012 & $0.287^{+0.019}_{-0.017}$&72/64&122/116[58/64]&72/64/0&201/$+$0.02&1.06&72/64/0&120/$-$0.03& 1.03& 879.4$\pm$5.4& 0.013\\
             &      &      &                          &best   &in reg &    &      &      &71/64/+0.2&88/$-$0.02& 0.98& & \\
             &      &      &                          &best   &global &    &      &      &82/71/$-$0.2&34/$-$0.03& 0.86& & \\
%             &      &      &                          & &&     &      &    &57/66&2840&0.59& & \\
J1952$+$3252 &36.57 &0.016 & 2.0$\pm0.5$              &\tablenotemark{b}&\tablenotemark{b} &71/84/0&19/$+$0.05&1.10&66/78/0&14/$+$0.05&0.84&  13.4$\pm$0.9& 0.015\\
             &      &      &                          &best   &in reg &  86/66/-0.2  &18/$+$0.05&1.10  &71/71/$-$0.2&14/$+$0.05& 0.70& & \\
J1709$-$4429 &36.53 &0.017 & 1.4$-$3.6                &34/53 &108/127[72/53]&31/49/0& 147/$+$0.10&1.30&36/56/0& 30/$-$0.04&0.89&   124$\pm$2.6&  0.24\\
             &      &      &                          &best   &global &    &      &      &47/53/+0.2&19/$-$0.01& 0.76& & \\
Geminga      &34.52 &0.175 & $0.25^{+0.12}_{-0.06}$   &--  &--&50/26/0&258/$+$0.41&0.93&21/89/0&241/$+$0.87&0.13& 338.1$\pm$3.5& 0.05\\
 & & & & & & & & & & & & & \\
J1057$-$5226 &34.48 &0.183 &  0.71$\pm0.2$            &  &105/111[75/69]&75/69/0&414/$+$0.12  &0.91 &75/69/0 &106/$+$0.11&0.82&  27.2$\pm$1.0& 0.45\\
             &      &      &                          &best   &global &41/08/-0.2&37/$+$0.03  &0.38 &\,7/87/-0.2&\,5/$+$0.11& 0.09& & \\
%J2021$+$3651 & 36.5 & 86(79)&     & 0.465 & 0.183 & 55-90 & 79 & 0.9-1.25 & 60-90 & 79 & 0.9-1.1\\
\enddata
\tablenotetext{a}{From \citet{psrcat}, flux units $10^{-11}{\rm erg\,cm^{2} s^{-1}}$}
\tablenotetext{b}{ ${\rm d}\Psi/{\rm d}\phi|_{max} \approx +3.5 {\rm ^\circ/deg}$ -- Weisberg et al. (1999).}
%\tablenotetext{c}{$\alpha=123^\circ$, rectified.}
%\tablenotetext{d}{$E_\gamma>0.1$Ge fluxV, units $10^{-11}{\rm erg\,cm^{2} s^{-1}}$}
\end{deluxetable*}

\subsection{Luminosities}

	Now that we have estimated the corrections to the all-sky flux $f_\Omega$,
it is worth checking how well the pulsars agree with the simple luminosity law 
Equation (1). A major challenge to this check is the distance imprecision;
of the five pulsars considered here only Vela and Geminga have parallax
distance estimates. The other distances (derived from Dispersion Measure DM) 
are subject to appreciable uncertainty.

Geminga is in reasonable agreement with the rule only if it lies at the upper
end of the present distance range. PSR J1057$-$5226 is nominally somewhat more 
luminous than expected,
but can be accommodated if the DM distance is a slight overestimate. 
PSR J1709$-$4429, on the other hand, is much brighter than expected unless
the true distance is $\sim 0.7$\,kpc or the flux correction factor is
as small as $f_\Omega \sim 0.07$ (or some combination of the two factors). Such 
changes would be quite surprising. It would be very interesting to obtain a 
parallax distance estimate for PSR J1709$-$4429 to eliminate that source
of uncertainty. 

\section{Conclusions and Future Prospects}

	In \citep{wet09} we showed that patterns in the separation of radio and 
$\gamma$-ray pulse components could be used to find acceptable geometrical
parameters in dipole magnetosphere models for pulsar emission. Here we show
that with high quality LAT light curves and reference to multi-wavelength
information on the inclination and viewing angles the direct comparison between the model
light curve shape is even more constraining. Within the restricted context
of the present computations -- dipole fields, simple expressions for the gap ranges
and $\gamma$-ray emissivity on the radiating surface -- we can already make some
strong statements.

\smallskip\noindent$\bullet$ Emission starting above the null charge surface 
(OG model) is strongly statistically preferred over models which have 
substantial emission starting from the stellar surface (eg. the TPC model).
For PSR J1952+3252 the fit for the TPC model is not much worse than that for 
OG, but for none of the pulsars modeled here is it better, if external 
angle constraints are used.

\smallskip\noindent$\bullet$ More realistic dipole field geometries, including
sweep back and fields enforcing co-rotation produce improved matches to
the pulsar light curves when one focuses on solutions near the orientation 
angles $\alpha$ and $\zeta$ implied by lower energy data. Some additional
improvement can be found by using test models for field-induced current
perturbations, but these are evidently not yet sufficiently realistic for detailed fits.

%\smallskip\noindent$\bullet$ The OG matches to certain peculiar light curves are
%quite good and are often found only close to the externally determined
%angles. 

\smallskip\noindent$\bullet$ The correction from observed flux to 
pulsar luminosity can be substantial. With these corrections
a simple constant $\Phi$, $L_\gamma \propto {\dot E} (\propto $ current) 
law provides an improved description of the data. Nevertheless,
the large departure for PSR J1709$-$4429 suggests additional factors
are needed to understand pulsar luminosities.

\smallskip\noindent$\bullet$ The best-fit global models are often several degrees
off of the externally known $\zeta$ and $\alpha$. This difference is statistically
significant and implies that the LAT data have the power to reveal perturbations
to the simplified field geometry.
\bigskip

	The robust success of the pulse profile computations described here 
for angles near those available from other data make it likely that the 
true magnetospheric geometry is fairly close to our simplified PFF model. 
It must be re-emphasized that this PFF model is not a complete physical
description -- and indeed the whole electrodynamic picture of `Holloway-type'
outer gaps is still of questionable validity. However, other scenarios
need to match the light curves quite closely to be competitive --
this likely forces them into rather similar magnetosphere geometries.
Comparison with other approaches, such as numerical magnetosphere simulations, 
should be fruitful. With more pulsar light curves and higher quality data
being provided by the {\it Fermi} LAT, we should be able to use the departures
from the predictions of these simple beaming laws to infer
geometrical modifications (and work toward their physical origin) and
to follow the evolution of the gap geometry and radiation with pulsar
age. In a following paper we describe the additional constraints
on $w({\dot E})$, $\delta w$, 
the radial extent of the emission zone, and the pulsar evolution with age
that can be extracted from the full LAT pulsar data set.
This should set the context for true radiation models, which can take
the next step in revealing the underlying magnetospheric physics.
\bigskip

	We thank Jon Arons, Sasha Tchekhovskoy and Anatoly Spitkovsky for useful
discussions about the field geometry. This work was supported in
part by NASA grants NAS5-00147 and NNX10AD11G.

\newpage
\appendix

For the convenience of other researchers we present here figures that
summarize the pulse properties of the TPC and OG models for the PFF dipole
geometry assumptions. These may be compared with the corresponding figures
in Watters et al. (2009).
\begin{figure}[h!!]
\vskip 21.2truecm
\includegraphics{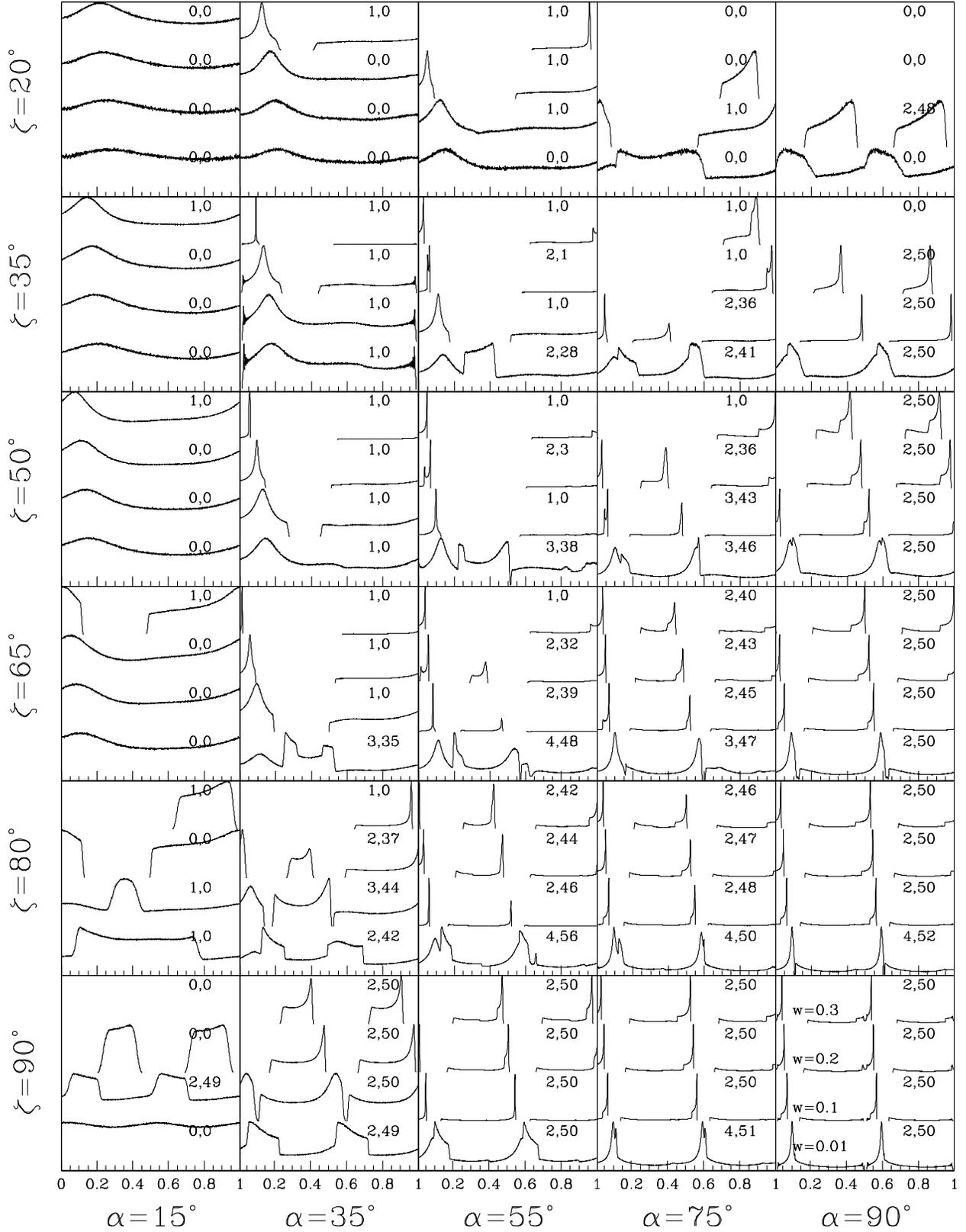}
\caption{\label{TPC_LC} Light curves for the Two Pole Caustic (TPC) model.
Each panel shows curves for four values of the gap width $w$. The curves
are labeled with the number of major peaks and the peak separation, in
percent.
}
\end{figure}

\begin{figure}[h!!]
\vskip 21.0truecm
\includegraphics{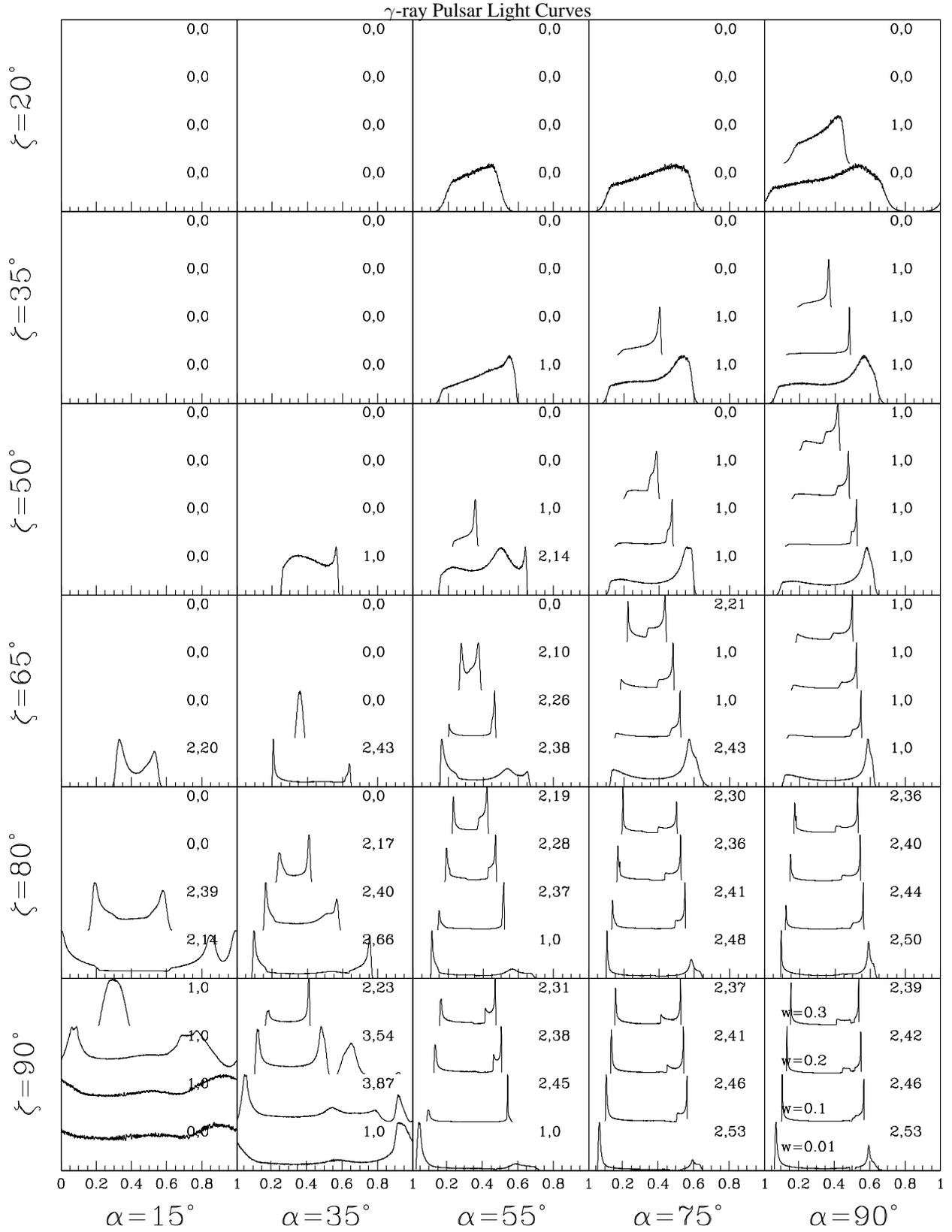}
\caption{\label{OG_LC} Outer Gap (OG) lightcurves. Labels as for Figure \ref{TPC_LC}.
}
\end{figure}

\begin{figure}[h!!]
\vskip 20.0truecm
\includegraphics{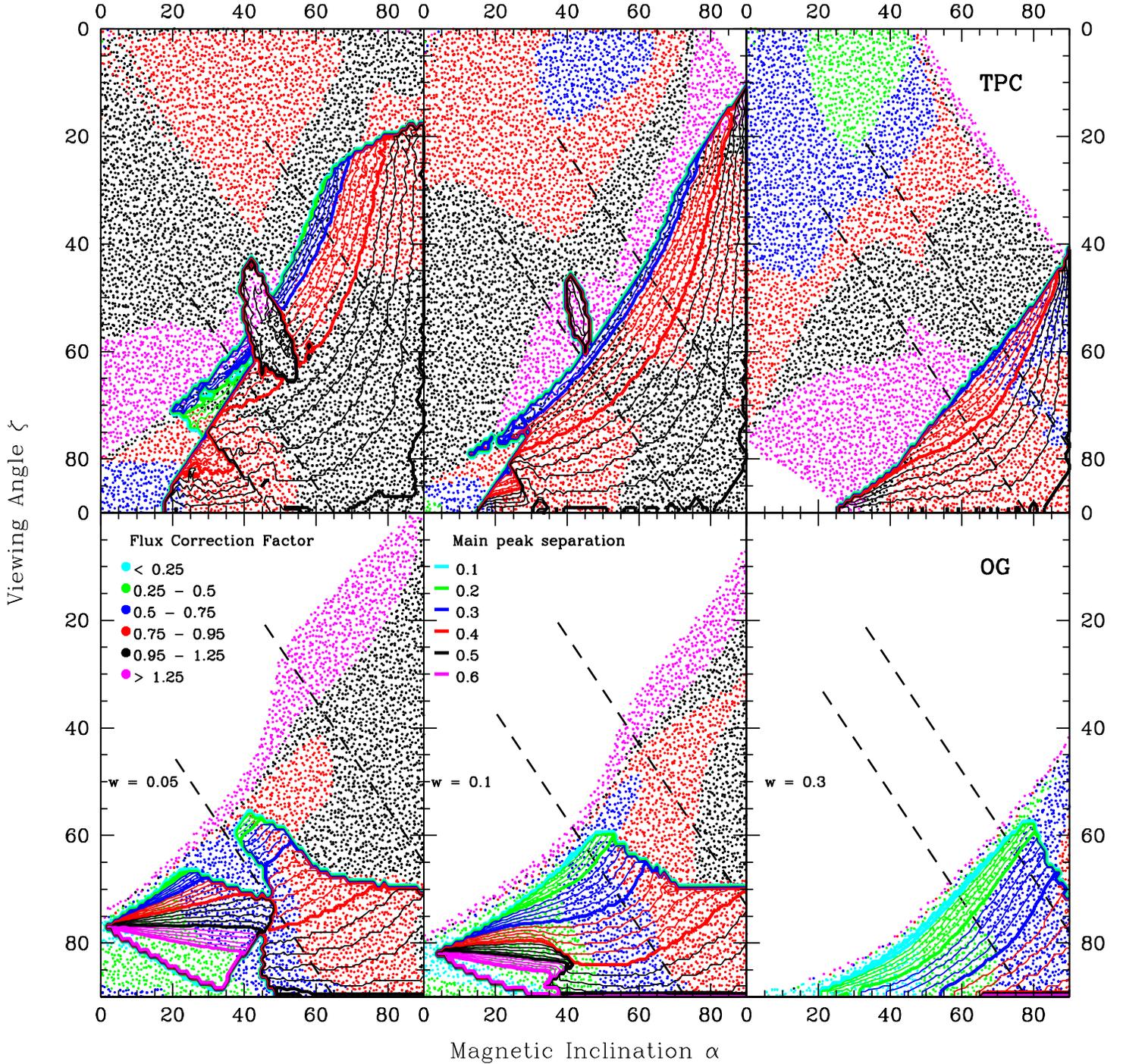}
\caption{\label{DeltafOm} Pulse properties in the $\alpha - \zeta$ plane for 
two pulsar emission geometries using the PFF field. Contours show the 
separation $\Delta$ of the principal $\gamma$-ray pulse peaks. Bold lines
mark $\Delta$ at 0.1 intervals (key bottom middle panel); fine lines mark 0.02 
intervals. The background shading (key bottom left panel)
gives the flux correction factor $f_\Omega$; large values mean that for the given
$\alpha$ and $\zeta$, the pulsar displays less flux than the sky average. 
White regions mean little or no flux from the modeled gaps.  The
dashed diagonal band in each panel indicates the region where lower altitude radio
pulsations are likely detected. Objects seen outside these bands tend to be Geminga-like
radio-quiet pulsars.
}
\end{figure}

\end{document}